\documentstyle[12pt,epsfig,subfigure,glas]{article}
\begin{document}
\begin{titlepage}{GLAS--PPE/95--06}{\today}
\title{\bf Noise Spectra of SIU-GaAs Pad Detectors With Guard Rings}
\begin{Authlist}
R.L.~Bates\Instref{phy_gla}
S.~D'Auria\Instref{phy_gla}
S.J.~Gowdy\Instref{phy_gla}
V.~O'Shea\Instref{phy_gla}
C.~Raine\Instref{phy_gla}
K.M.~Smith\Instref{phy_gla}
A.~Longoni\Instref{phy_mil}
G.~Bertuccio\Instref{phy_mil}
G.~De~Geronimo\Instref{phy_mil}
\end{Authlist}
\Instfoot{phy_gla}{Dept. of Physics \& Astronomy, University of Glasgow, UK}
\Instfoot{phy_mil}{Department of Physics, Politecnico di Milano, Milano,
Italy}
\collaboration{On Behalf of the RD8 Collaboration}

\begin{abstract}
This paper presents current noise characterization of circular pad Schottky
barrier diodes with guard rings. The diodes were
fabricated from undopped semi-insulating GaAs, SIU-GaAs, at the University of
Glasgow.
Current noise spectra were obtained for the detectors for two pad sizes, with
reverse bias applied. Three measurements were also made on one of the
detectors
under forward bias. The noise spectra show an excess noise component, with a low
frequency corner at less than 1kHz, and
a flat region at higher frequencies. The magnitude of the white noise is
approximately half that
expected from shot noise theory for the given leakage currents. 
A fall in the magnitude of the noise was observed at 20kHz which
is attributed to the dielectric relaxation time of the material. 

\end{abstract}

\vspace{2cm}
\centerline{\em Presented by R.L.~Bates at the $4^{th}$
workshop on GaAs detectors and
related compounds}
\centerline{\em San Miniato (Italy) 19-21 March 1995}
\end{titlepage}
  
\section{Noise theory}
A detailed discussion of noise 
theory is given in references 1 and
2. 

For any quantity {\em X(t)} that exhibits noise, the noise 
power within a unit bandwidth or power spectral density {\em S$_{X}$(f)} is
defined as 
\begin{equation}
\label{sdef}
S_{X}(f) = \lim_{T\to\infty} { {\overline{ 2|{\it X}(f)|^{2} }} \over{T} }  
\end{equation}
where {\it X(f)} is the Fourier transform of {\em X(t)},
and ${\overline{{|{\it X}(f)|}}}$ is an ensemble average.
In the simplest case where the transitions that cause the noise are 
described by equation (\ref{equan3}), where $\tau$ is the lifetime 
of the fluctuation causing interaction,
the spectral density is given by equation (\ref{equan4}).
\begin{equation}
\label{equan3}
{-d \Delta X\over dt} ={ \Delta X \over \tau}
\end{equation}
\begin{equation}
\label{equan4}
S_{X} (f) = {<(\Delta X)^{2}> 4\tau \over (1+(2\pi f \tau)^2)}
\end{equation}
From simple statistical considerations $<$($\Delta~X$)$^{2}>$ 
can usually be found, for example in the case of number fluctuations 
it is given by Poisson statistics.

This equation is very general with the condition imposed that the 
interactions of the electrons are independent. For small
fluctuations this is indeed true and thus the Lorentzian spectrum (equation
(\ref{equan4})) 
appears often. At 
low frequencies ($f\tau$~$<<$1) the spectrum is
white, that is independent of frequency, while at
high frequencies ($f\tau$~$>>$1) it varies as 1/$f^{2}$, and its half power 
point is at f=1/(2$\pi\tau$).

\subsection{Equivalent Noise Generators}

In a two terminal network, noise in a frequency interval $\Delta~f$ may be
characterized 
using either an equivalent e.m.f. generator
$\sqrt{S_{v}(f)}$ in series with the device or a current generator 
$\sqrt{S_{i}(f)}$ in parallel. An equivalent noise resistance,
$R_{n}$, and an equivalent noise current, $I_{eq}$, may be defined  
\begin{equation}
\label{noiseres}
S_{v}(f) = 4kTR_{n}
\ , \  S_{i}(f) = 2eI_{eq}
\end{equation}
where $T$ is the room temperature and $k$ is Boltzmann's constant.
\begin{figure}
\centerline{\epsfig{file=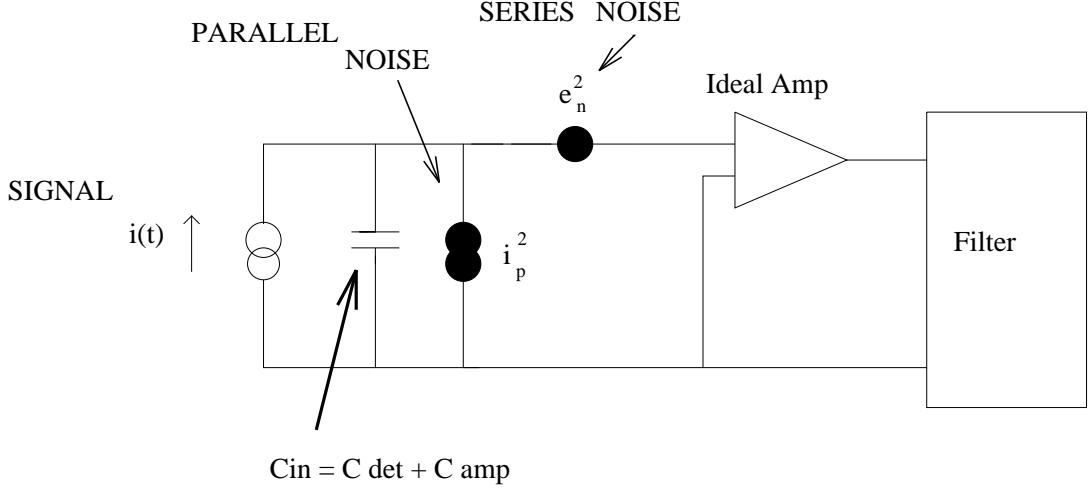,width=2.5in,angle=270}}
\caption{Equivalent circuit of the detector and amplifier for noise
considerations
      \label{eqcir}}
\end{figure}

The equivalent circuit for a detector and amplifier system is shown in figure
\ref{eqcir} 
and the noise components for an FET amplifier input stage are given below.
\begin{equation}
\label{serisenoi}
S_{v}(f) = N_{v}+ {A_{f}\over{f}}
\ , \  
N_{v} = {3\over{2}}{4kT\over{g_{m}}} 
\end{equation}
\begin{equation}
\label{parallelnoi}
S_{i}(f) = N_{p} + {B_{f}\over{f}}
\ , \  
N_{p} = 2e(I_{det}+I_{tran}) + {4kT\over{R_{f}}} 
\end{equation}
where $g_{m}$ is the transistor transconductance.
The equivalent noise charge of a circuit with shaping time $\tau$, is then 
\begin{equation}
\label{enc}
ENC^{2} = S_{1} \ {1\over{\tau}} \ {C_{in}}^{2} \ N_{v} \  
+ \ S_{2} \ \tau \ N_{p} 
+ \ {\em O}({1\over{f}}) \ 
\end{equation}
where $S_{1}$ and $S_{2}$ depend upon the type of shaping used.
Thus an an optimum shaping time exists.
If $S_{1}$ and $S_{2}$ are of similar magnitude and the leakage current term
dominates, 
a short shaping time is required. The
minimum obtainable equivalent noise charge may be limited by the {\em 1/f} noise
if this term is large.

\subsection{Types of noise sources}

There are four noise source classifications in semiconductors: thermal, shot, 
generation-recom\-bination, and modulation noise. 
The first three are well understood
while the origin of the fourth with regard to semiconductors is less well so.

Thermal noise is a white noise source whose origin is based on fundamental
thermodynamic physical laws.
For a semiconductor of resistance {\em R} the 
spectral current and voltage noise densities are
\begin{equation}
\label{equan5}
S_{i}={4kT\over{R}} \ \ \  S_{v}={4kTR}
\end{equation}
 
Shot noise is due to the discreteness of the charge carriers
and is related to the statistical nature of their injection into 
the semiconductor over a Schottky barrier. 
The origin of the spectral
density may be found by considering Carson's Theorem \cite{Ziel}. 
This states that for a diode with reverse current {\em I} which equals $e<n>$,
where {\em n} is the spontaneously 
fluctuating number of electrons that cross the barrier per second, the spectral
density 
is given by 
\begin{equation}
\label{equan6}
S_{i}(f) = 2e^{2}<n> = 2eI 
\end{equation} 

It can be seen that the shot noise is linearly dependent on the frequency of the
electrons crossing the barrier and
quadratically dependent on the charge of the pulse that the electron creates in
an external circuit. 
The spectrum is white because of the very short transit time of the carriers
across the barrier, where the barrier includes
the space charge region.
For the barrier to show full shot noise at all frequencies two conditions must
be met. 
The first is that when charge carriers are
injected into the
material, space-charge neutrality is re-established in a very short time, given
by the material dielectric relaxation
time $\epsilon\rho$ \cite{sze}, where $\rho$ is the resistivity of the material.
For SIU-GaAs this is of the order of $10^{-5}$s.
The second is that each current pulse should be able to displace a charge
equivalent to that of one electron in the external
circuit, that is to
say a low trap density is required at the metal-semiconductor interface. 

Generation-recombination noise can be understood by considering a semiconductor
with a number of traps. 
The continuous trapping and de-trapping of the charge carriers causes a
fluctuation
in the number of carriers in the conduction and valence bands. The transitions
are described by equation (\ref{equan3}) and thus the
noise spectral density is Lorentzian in nature, with a corner frequency given by
$\tau _{n}$, the lifetime of the electrons in the
conduction band. The current noise density was calculated by van Vliet
\cite{boer} to be
\begin{equation}
\label{equan7}
{S_{n} (f) ={ <(\Delta n)^{2}> 4 \tau_{n} \over (1+(2\pi f \tau_{n})^2)} }
\end{equation} 
 where $<(\Delta n)^{2}> \propto I^{2} $.
It should be noted that in observed
spectra the corner frequency of this noise is dispersed, implying a wide
distribution of lifetimes. If the current that flows
through a device is due to generation then the expression for the noise present
will be the generation-recombination one rather
than the expression for the shot noise defined in equation (\ref{equan6}).
However in most Schottky junction devices the applied
voltage will increase the injection rate and the noise will show a very close
resemblance to shot noise.

At low frequencies an excess noise spectral density with a {\em 1/f} amplitude
dependence 
is observed in semiconductors, which is known as modulation noise.
As stated earlier the exact cause of this noise is still not understood.

\section{The Diodes}
The material used for the detectors was 200$\mu$m thick semi-insulating GaAs.
The detector fabrication was performed at the 
University of Glasgow. The
detectors were designed with a circular pad contact and guard ring on top of the
substrate. The bottom
contact was a uniform contact that spread to the edge of the substrate.
Two geometries were used, one with a pad diameter of 2mm the other of 3mm; the
width of the guard was 200$\mu$m and
the pad guard 
separation was 10$\mu$m for both diameters. The pad and guard metallization
layers were identical and rectifying in nature. The
reverse contact had a different recipe but was also
rectifying,
however its saturation current was larger than the pad's due to its larger
surface area. In the following, the direction of the
detector bias is quoted with respect to that of the pad. Thus reverse bias means
that the
pad is at a negative potential with respect to the back contact.

\begin{figure}
\begin{center}
\begin{tabular}{c}
\subfigure[Forward Bias]{\epsfig{file=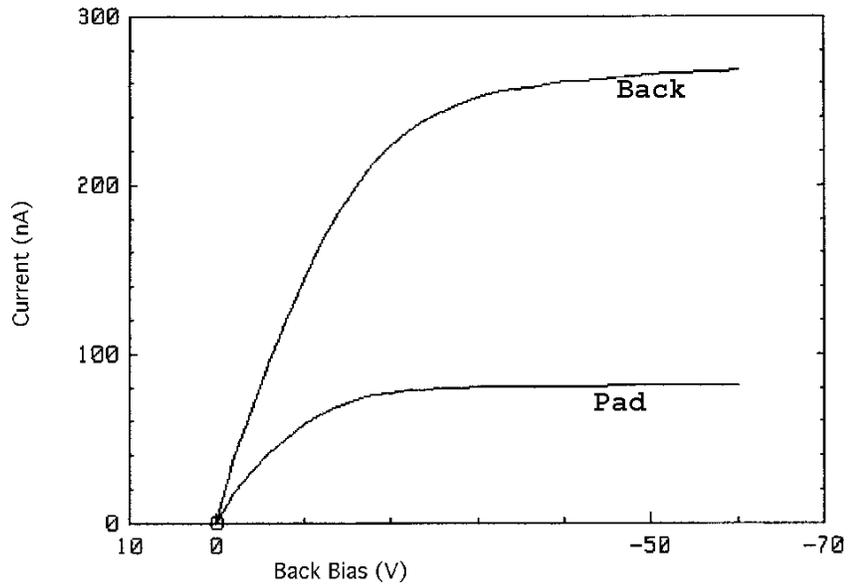,height=3in}} \\
\subfigure[Reverse Bias]{\epsfig{file=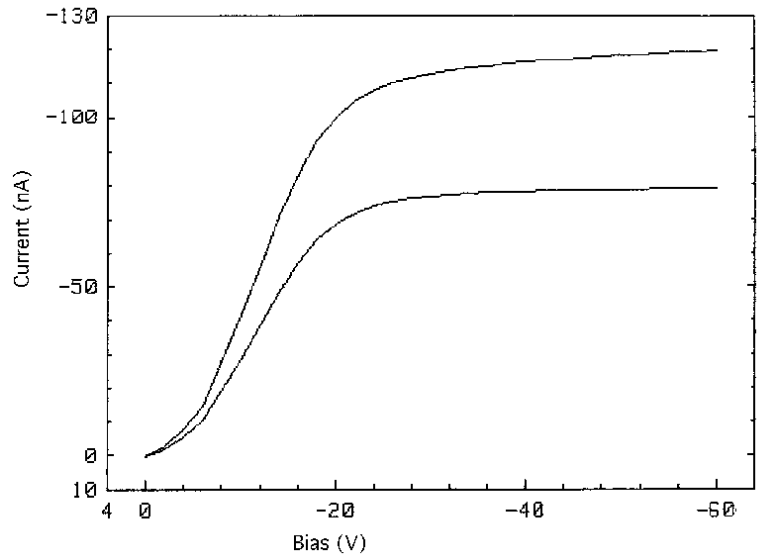,height=3in}}
\end{tabular}
\end{center}
\caption{Current-Voltage characteristics, Temp=300K\label{figiv}}
\end{figure}

The {\em I-V} curves obtained for the diodes in reverse and forward bias are
shown in figure \ref{figiv}.
The purpose of the guard ring 
was to ensure that the current through the pad remained constant once in
saturation.
This was observed for all the samples, with the 
current that flowed through the pad, in the bias range 60
to 200 volts, increasing by only a few percent.
For a given bias the current densities of both pads were the same and equalled
25nA/mm$^{2}$.
The results showed that the saturation
current density that flowed through the pad under forward bias was the same as
that obtained in reverse bias.

\section{Measurement Procedure}
To measure the current noise of the diodes under an applied bias an amplifier
designed by the Politecnico di Milano was used
\cite{milano}. This had a large bandwidth ($\omega_{-3dB}\simeq 20kHz$), high
gain ($\simeq 86dB$), low noise
($\sqrt{\overline{v_{ni}^{2}}}\simeq 1.
8nV / \sqrt{Hz} @ 1kHz $) and an adjustable input bias offset. The amplifier had
a large input impedance ($>$1G$\Omega$)
and a low output impedance ($<150\Omega$). This amplifier was used as an
operational amplifier with a feedback resistor $R_{f}$ and
a feedback capacitance $C_{f}$. The output of the amplifier was passed through
the high impedance input of an AC coupled
oscilloscope and out through the 'scope's $50\Omega$ output. The 'scope thus
acted as a low noise voltage amplifier. The output of
the 'scope was sent to a HP signal analyser with $50\Omega$ input impedance.
The feedback resistor was chosen to maximize the gain of the amplifier (large
$R_{f}$), 
minimize the background noise due to the thermal noise of the feedback 
resistor (large $R_{f}$, see equation (\ref{equan5})) and to make sure that the
amplifier did not saturate (limits $R_{f}$). 

Before the current noise density spectra of the diodes were measured the
transfer function 
and the background noise of the amplifier were 
determined. The transfer function was required so that the measured voltage
noise 
could be referred back to the input as a current noise.
Background noise measurements were made to show that the noise contribution from
the
'scope was negligible and that the overall background noise 
present in normal operation was close to the thermal noise of the feedback
resistor.

Noise spectra were obtained for both diodes, at zero bias, under reverse bias,
at the same 
leakage current for both diodes and under forward bias for the small diode at
three leakage
currents equivalent to those used for the reverse bias case. 

\section{Results}
In this section the current noise spectral densities of the diodes are shown 
and some tentative interpretations of the results are proposed.

At zero bias the measured noise is the thermal
noise of both the feedback resistor and the diode. From equation
(\ref{noiseres}) the 
equivalent noise resistances of the diodes
were found which with the use of the diode dimensions enabled resistivities for
the material
of 12.4M$\Omega$cm and 18.3M$\Omega$cm to be obtained. Both values are of the
correct order of magnitude.

\begin{figure}
\centerline{\epsfig{file=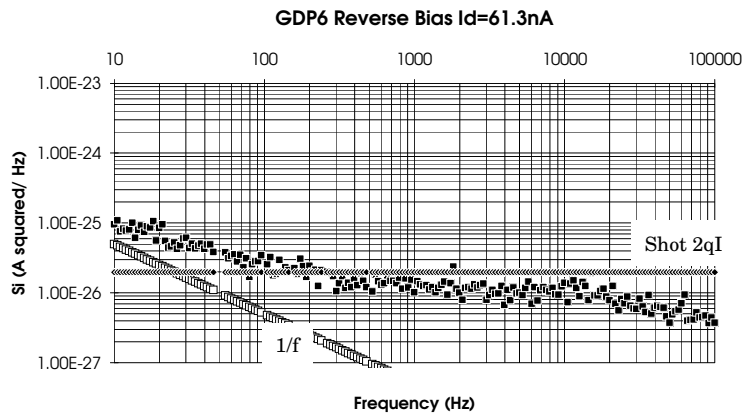,height=2.5in,angle=0}}
\caption{Typical noise spectrum of the small GaAs diode under reverse bias
      \label{61}}
\end{figure}

Figure \ref{61} shows a typical noise spectrum. All the spectra are white at
frequencies above the 
corner frequency of the excess noise. The value of the current noise
density is less that that expected from equation (\ref{equan6}). At a frequency
of 10kHz the 
noise over the measured leakage current range is between 1.7 and 2.3 times less
than the simple shot noise theory. At frequencies between 20kHz and 30kHz there
is a corner in the 
white noise.

\begin{figure}
\centerline{\epsfig{file=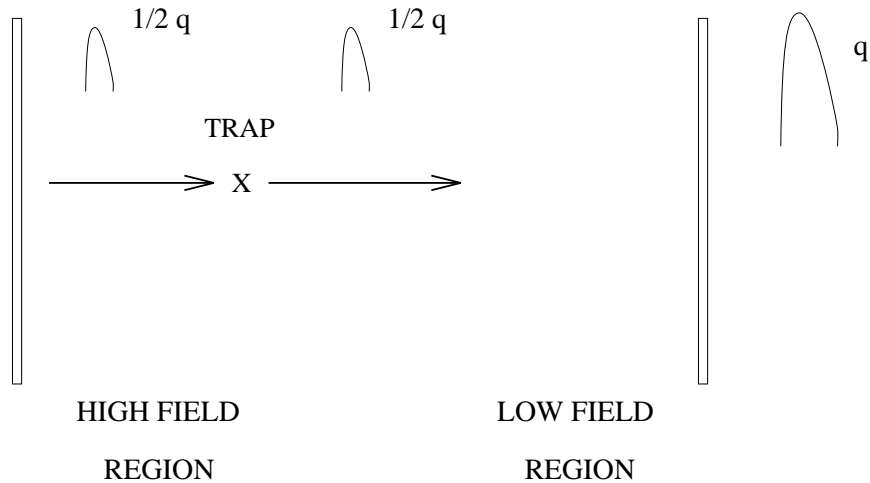,width=2.5in,angle=270}}
\caption{Simple model for shot noise.
        \label{simpshot}}
\end{figure}

If the leakage current is due to the thermionic emission of carriers over the 
Schottky barrier then the noise should be shot noise. A simple model can be
devised 
to explain the shot noise spectra obtained, with reference to equation
(\ref{equan6}) and figure \ref{simpshot}. 
One can say that if an
electron is trapped and at a later instant released then this electron will
produce two 
pulses of charge whose sum is equal to {\em e}. If
the electron is captured after producing a charge pulse equal to ${1\over{2}}~e
$ in the external 
circuit and then released, on average
the frequency of charge pulses produced by the emitted electrons is doubled
while the charge is halved. The shot
noise spectral density is thus reduced by a factor of two. 
Shot noise should be white due to the fast transit time of the electrons across
the high field region behind the reversed biased
Schottky junction, however, this is not observed. The corner frequency at 30kHz
can be explained with reference to the condition imposed that 
charge neutrality is obtained in a time
given by the dielectric relaxation time. Assuming that the noise source is
Lorentzian, the corner frequency
corresponds to the reciprocal of this time constant. Thus a lifetime of the
order of
10$^{-5}$s is required, which is equivalent to the dielectric relaxation time in
SIU-GaAs.

If the major contribution to the reverse bias leakage current is the generation
current, rather than the injection of carriers over
the Schottky barrier then the noise is due to
generation-recombination noise from the bulk and not thermionic emission at the
barrier. The noise due to generation has a
characteristic time constant, the lifetime of the generated electron in the
conduction band, which is inversely proportional to the corner frequency of the
spectrum. 
If the transport mechanism is said to be by relaxation processes then the
lifetime, $\tau_{d}$ in equation 
(\ref{equan7}), is the dielectric relaxation time.
Although, in principle, the dependence of the measured spectral density on the
leakage current should help separate the two causes of the white noise,
no clear dependence was observed.

The final part of the diode noise spectrum is the low frequency excess noise.
This is seen for measurements made on the small diode for currents up to 70nA
(75V bias) as approximately
{\em 1/f} noise which meets the white noise at a frequency of 500Hz. At 10Hz the
excess noise
is an order of magnitude greater than
the white noise component of the spectrum. 
This is also observed in the large diode for currents up to 170nA (50V bias).

\begin{figure}
\centerline{\epsfig{file=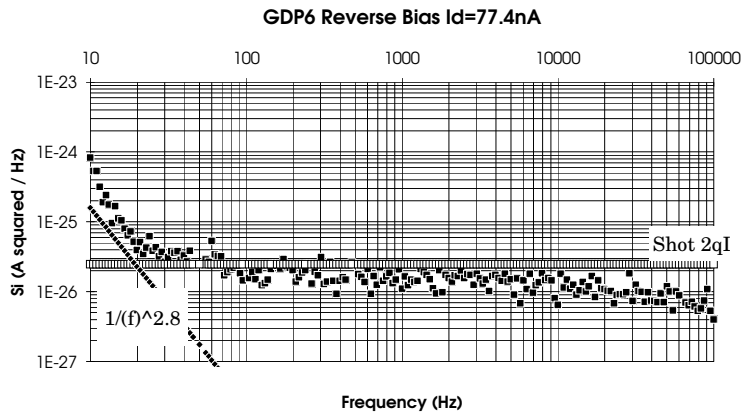,height=2.5in,angle=0}}
\caption{Spectrum showing the steep frequency dependence of the excess noise.
      \label{steepnoi}}
\end{figure}

At larger currents (above 71nA for the small diode) a second low frequency 
noise component appears which has a much steeper
frequency dependence, $1/f^{\alpha}$ where $\alpha$ is close to 3. 
At 10Hz the excess noise is now almost two orders of magnitude larger than the
white noise.
(figure \ref{steepnoi}).

\begin{figure}
\centerline{\epsfig{file=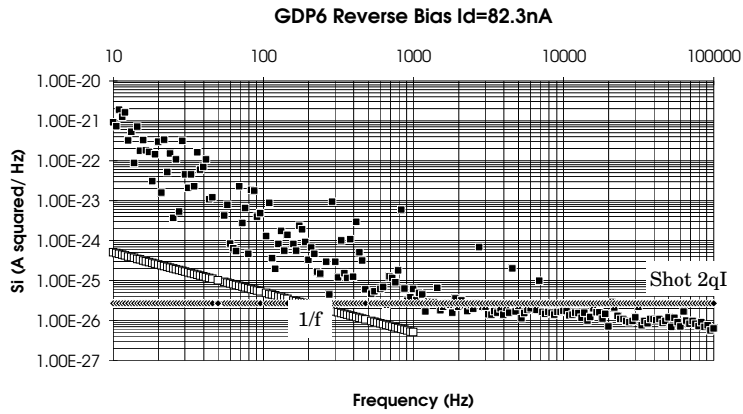,height=2.5in,angle=0}} 
\caption{Noise spectrum of the small GaAs diode at a leakage current of 82nA,
reverse bias of 230V.
      \label{excess1}}
\end{figure}

\begin{figure}
\centerline{\epsfig{file=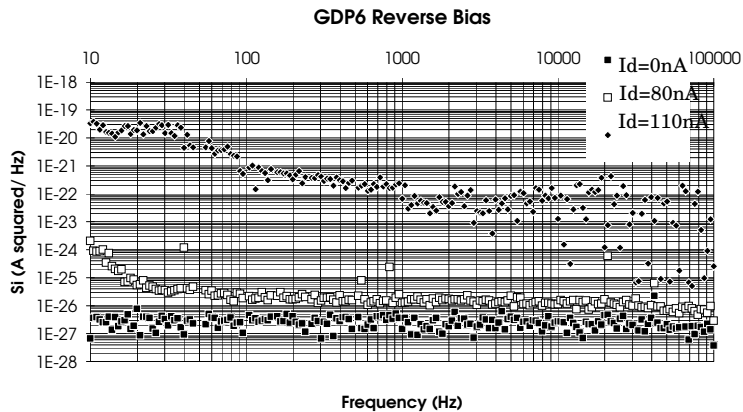,height=2.5in,angle=0}}
\caption{Noise spectra of the small GaAs diode for leakage currents of 0, 80 and
110nA.
      \label{excess2}}
\end{figure}

The magnitude of the excess noise is seen to increase with bias towards the
`breakdown' of the diode. Here the corner frequency of the excess noise extends
to much higher frequencies; 1kHz @ 230V, and
$>$100kHz @ 240V for the small diode. The magnitude of the noise increases
dramatically, to reach 3$\times$10$^{-21}$ A$^{2}$/Hz at 10Hz
at 230V for the small diode (see figures \ref{excess1} and \ref{excess2}). At a
bias of 200V the electric field extends
across the sample to the forward biased contact
which causes charge injection and thus the dramatic increase in the measured
noise which occurred between 230V and
240V.

\begin{figure}
\centerline{\epsfig{file=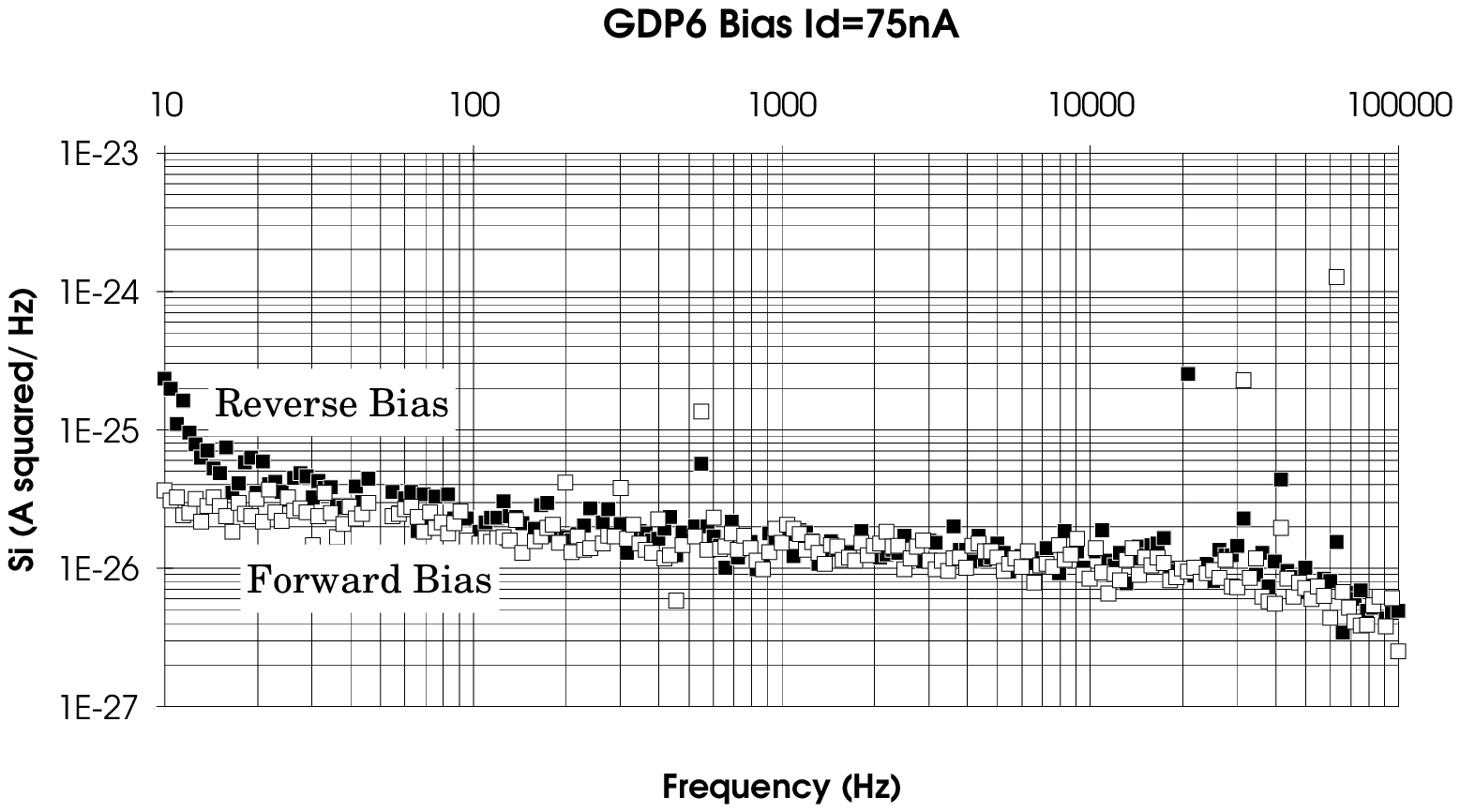,height=2.5in,angle=0}}
\caption{Noise spectra of the small GaAs diode in reverse and forward bias.
      \label{figrf75}}
\end{figure}

Noise spectra for the forward biased diode are similar to those obtained in
reverse bias 
as shown in figure \ref{figrf75}. The white noise component is
approximately the same, the high frequency corner is present, but no low 
frequency excess noise is present in the forward biased
case.
The reason why
no excess noise is seen for the forward biased diode is not clear. 

\section{Conclusions and Future Work}
The noise measurements show three distinct regions in the spectra, the first
being the low frequency, $1/f$, excess noise
which has a corner frequency of 500Hz for bias voltages below 75V. Above this
the excess noise has a second component which is
steeper,
being almost $1/f^{3}$ in nature, and has a corner frequency of 50Hz which
increase with bias. As the diode approaches breakdown
the corner frequency and the magnitude of the excess noise increases
dramatically due to injection at the forward biased contact.
The second feature of the spectra is the lower than expected value of the shot
noise, being about half that of simple theory. This
is due to trapping of the charge carriers in the bulk which reduces the
amplitude of the individual pulses that the carriers
produce while increasing their frequency. The third region is the corner
frequency of the white noise between 20-30kHz. This is
noted to be close to the reciprocal of the dielectric relaxation time for
SIU-GaAs. 

Further noise measurements need to be made on more samples using different types
of contacts to see if there is any 
correlation between the noise of the sample and the properties of the material
and fabrication. Diodes that have been
irradiated also need to be tested to improve the present understanding of the
effects of irradiation. The technique can also be
extended to examine strip detectors made from GaAs. The biasing structure used
at present utilizes a punch-through mechanism via
a MSM diode. There is a possibility that this may introduce more noise and it is
important to understand if this is
true.

\section{Acknowledgements}
The authors wish to express their appreciation to
A.~Meikle, F.~Doherty and F.~McDevitt of the University of Glasgow for their
fabrication skills.

\end{document}